\documentclass[a4paper]{jpconf}
\usepackage{graphicx}
\usepackage{float}
\usepackage{wrapfig}

\usepackage{bm}
\usepackage{amsmath,amssymb}
\usepackage{citesort}
\usepackage{threeparttable}
\usepackage[dvipsnames]{xcolor}

\usepackage[normalem]{ulem}  

\newcommand{\HI}{H\,{\sc i}}

\newcommand{\HeII}{He\,{\sc ii}}

\usepackage[textwidth=1in]{todonotes}

\def\apj{Astroph. J.}
\def\apjl{Astroph. J. Lett.}
\def\apjs{Astroph. J. Suppl.}
\def\mnras{MNRAS}

\def\aap{Astron. Astroph.}

\def\aj{Astron. J.}
\def\pasp{Pub. Astron. Soc. Pacific}

\begin{document}
\title{Jeans smoothing of the Ly$\alpha$ forest absorption lines}

\author{K N Telikova, S A Balashev and
P S Shternin}

\address{Ioffe Institute, 26 Politeknicheskaya st., St.\ Petersburg, 194021, Russia}

\ead{telikova.astro@mail.ioffe.ru}

\begin{abstract}
We investigate a contribution of the Jeans smoothing to the minimal width of Ly$\alpha$ forest lines and discuss how the accounting for this additional broadening affects the inferred parameters of the intergalactic matter equation of state. We estimate a power-law index $\gamma$ of the equation of state, a temperature at the mean density $T_0$ and a hydrogen photoionization rate $\Gamma$ within 4 redshift bins. Furthermore, in each bin we obtain an upper limit on the scale-parameter $f_{\rm J}$, which sets the relation between the Jeans length and the characteristic physical size of the absorber clouds. 
\end{abstract}

\section{Introduction}

It is believed that the effective equation of state (EOS) of the low-density intergalactic medium (IGM) after the \HI\ reionization obeys the power law \cite{HuiGnedin1997MNRAS}:
\begin{equation}\label{eq:eos}
  T= T_0 \Delta^{\gamma-1},
\end{equation}
where $T=T(\rho)$ is a temperature at a density $\rho$, $T_0 \equiv T(\bar{\rho})$ is the temperature at the mean density $\bar{\rho}$ of the Universe 
and $\Delta\equiv\rho/\bar{\rho}$ is an overdensity.
Evolution of the EOS, defined by a  dependence of $T_0$ and a power-law parameter, $\gamma$, 
on a redshift, is determined by the dynamics of the reionization processes. One of the widely used methods to probe the EOS is the statistical analysis of the parameters of the Ly$\alpha$ forest lines observed in spectra of distant quasars \cite{Schaye1999,Schaye2000,Rudie2012,Hiss2018}. This method exploits an assumption that the minimal broadening of the Ly$\alpha$ lines is due to thermal motions of the absorbing atoms. However, it was suggested that the Hubble expansion during the time that light crosses an absorber may result in the minimal line broadening that depends not only on the thermal velocity distribution within the absorber, but also on its spatial structure \cite{HuiGnedin1997ApJ,Garzilli2015,Garzilli2018arXiv}. In other words, an observed broadening of the Ly$\alpha$ lines 
encodes an information about the physical extent of the absorbers. In the present study we estimate a significance of the additional broadening related to the Hubble expansion and obtain 
constraints on the EOS parameters and the scale parameter between the Jeans length and the characteristic physical size of low density IGM absorbing clouds from the analysis of the observed joint distribution of column densities $N$ and Doppler parameters $b$ of Ly$\alpha$ forest absorbers.

\section{Data and method}

\begin{table}[t!]
    \centering
    \caption{Fit results for the model parameters. Parameters uncertainties correspond to the 68 per cent credible intervals. Here $\bar{z}$ is a mean redshift in a bin, $T_{04}$ is the temperature at the mean density, $T_0$, in units of $10^4$~K and other parameters are described in the text.}\label{tab:pars} 
    \begin{tabular}{cccccccccc}
        \br
		Redshift range & $\bar{z}$ & $\beta$  & $T_{04}$ & $\Gamma_{-12}$ &$\gamma-1$ & ${\gamma-1}^\dagger$\\ 
		\mr

		$1.90 - 2.38$& $2.23$ & $-1.73^{+0.09}_{-0.07}$ & $1.37^{+0.12}_{-0.19}$ & $0.86^{+0.19}_{-0.14}$& $0.49^{+0.06}_{-0.05}$&$0.47^{+0.04}_{-0.04}$\\

		\rule{0pt}{2.6ex}  

        $2.38-2.62$& $2.51$& $-1.55^{+0.08}_{-0.07}$  &$1.96^{+0.15}_{-0.93}$ & $0.66^{+0.34}_{-0.10}$ &$0.42^{+0.21}_{-0.09}$&$0.38^{+0.05}_{-0.06}$\\
		
	    \rule{0pt}{2.6ex} 

		$2.62-2.95$ & $2.78$ &  $-1.21^{+0.08}_{-0.08}$ & $1.98^{+0.17}_{-0.57}$ & $0.64^{+0.18}_{-0.09}$& $0.56^{+0.14}_{-0.06}$& $0.54^{+0.03}_{-0.05}$\\
		\rule{0pt}{2.6ex}

		$2.95-3.73$& $3.18$& $-1.18^{+0.07}_{-0.07}$ & $1.93^{+0.19}_{-0.77}$ & $0.70^{+0.33}_{-0.12}$& $0.33^{+0.24}_{-0.12}$&$0.27^{+0.07}_{-0.08}$ \\
		\br
    \end{tabular}
     \begin{tablenotes}
        \item $\dagger$ neglecting Jeans smoothing contribution
    \end{tablenotes}
\end{table}


\begin{figure}[ht]
\includegraphics [width=0.65\textwidth, trim=0 15 0 0, clip]{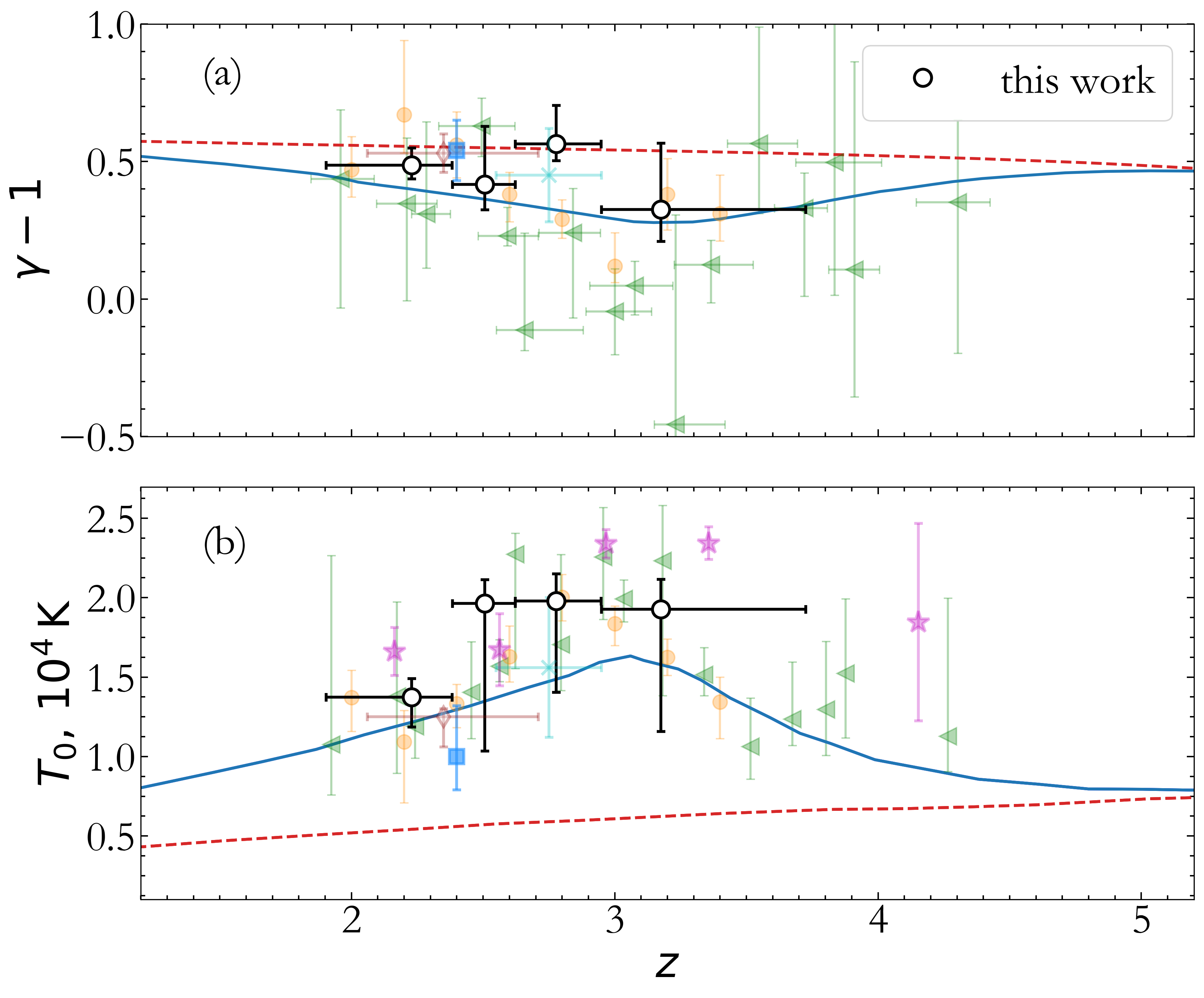}\hspace{1.5pc}%
\begin{minipage}[b]{12pc}\caption{\label{fig:gamma_T0_4}EOS parameters $\gamma$ (top panel) and $T_0$ (bottom panel) as functions of the redshift. Our results are shown by the black open circles. Blue solid and red dashed curves correspond to two different scenarios of the reionization from \cite{UptonSanderbeck2016}, namely, with and without accounting for the \HeII\ reionization, respectively. We also show measurements from \cite{Schaye2000} (green triangles), \cite{Bolton2014} (blue filled squares), \cite{Hiss2018} (yellow filled circles), \cite{Rorai2018} (the cyan cross), \cite{Telikova2018a} (the brown diamond), \cite{Lidz2010} (magenta stars).}
\end{minipage}
\end{figure}

We obtained a large sample of Ly$\alpha$ forest lines in the redshift range $z\sim2-4$ in 47 high-resolution ($R\sim36000-72000$) and  high signal-to-noise ratio ($S/N\sim20-100$) quasar spectra from KODIAQ\footnote{Keck Observatory Database of Ionized Absorption toward Quasars} \cite{OMeara2017} using the original fitting procedure, see \cite{Telikova2018a,Telikova2018b,Telikova2019} for details.

For the analysis of the obtained sample we employed the method developed in our previous works \cite{Telikova2018a,Telikova2018b,Telikova2019}. The method is based on the approximation of the $(N,\,b)$ sample  by the model probability density function
\begin{equation}\label{eq:pdf_N_b}
  f(N,\, b)=\int\limits_0^\infty f_N(N)f_{\rm add}(b_{\rm{add}})\delta \left(b-\sqrt{b_{\rm{min}}^2+b_{\rm{add}}^2} \right)\rm{d} \emph{b}_{\rm{add}},
\end{equation}
where $b_\mathrm{min}$ is the minimal broadening at a given $N$, $b_{\rm add}$ is an additional broadening, accounting for the turbulent and peculiar motions, $f_N(N)$ and $f_{\rm add}(b_{\rm{add}})$ are distribution functions of Ly$\alpha$ absorbers over $N$ and $b_{\rm add}$, respectively. It is well established, that $f_N(N)$ has a power-law shape, $f_N(N)\propto N^{-\beta}$ (e.g. \cite{Janknecht2006J,Rudie2013}). For $f_{\rm add}(b_{\rm{add}})$ we also assumed a power-law behaviour $\propto b_{\rm{add}}^p$ (for the discussion of this choice, see \cite{Telikova2018b,Telikova2019}).
Usually one suggests that the minimal broadening of the absorption lines is determined predominantly by the thermal contribution. Here we investigate the model proposed by Garzilli \etal\ \cite{Garzilli2015,Garzilli2018arXiv}, where the minimal broadening of the Ly$\alpha$ lines is a sum of two contributions
\begin{equation}\label{eq:bmin}
    b_{\rm min}^2 = b_{\rm th}^2 + b_{\rho}^2 \equiv \frac{2k_{\rm B}T}{m} + f_{\rm J}^2\left(\frac{\lambda_{\rm J}H(z)}{2\pi}\right)^2,
\end{equation}
where $k_{\rm B}$ is the Boltzmann constant, $m$ is the hydrogen atom mass, $\lambda_{\rm J}$ is the Jeans length and $H(z)$ is the Hubble constant. The parameter $f_{\rm J}$ introduced in \cite{Garzilli2015} describes the relation between the characteristic physical extent of the baryonic matter of IGM cloud and the Jeans length. The second term in eq~(\ref{eq:bmin}), $b_\rho$, referred as the broadening due to the Jeans smoothing,
results from a spatial structure of the absorber.
\begin{figure}[t!]
\includegraphics [width=0.61\textwidth,  trim=0 15 0 0, clip]{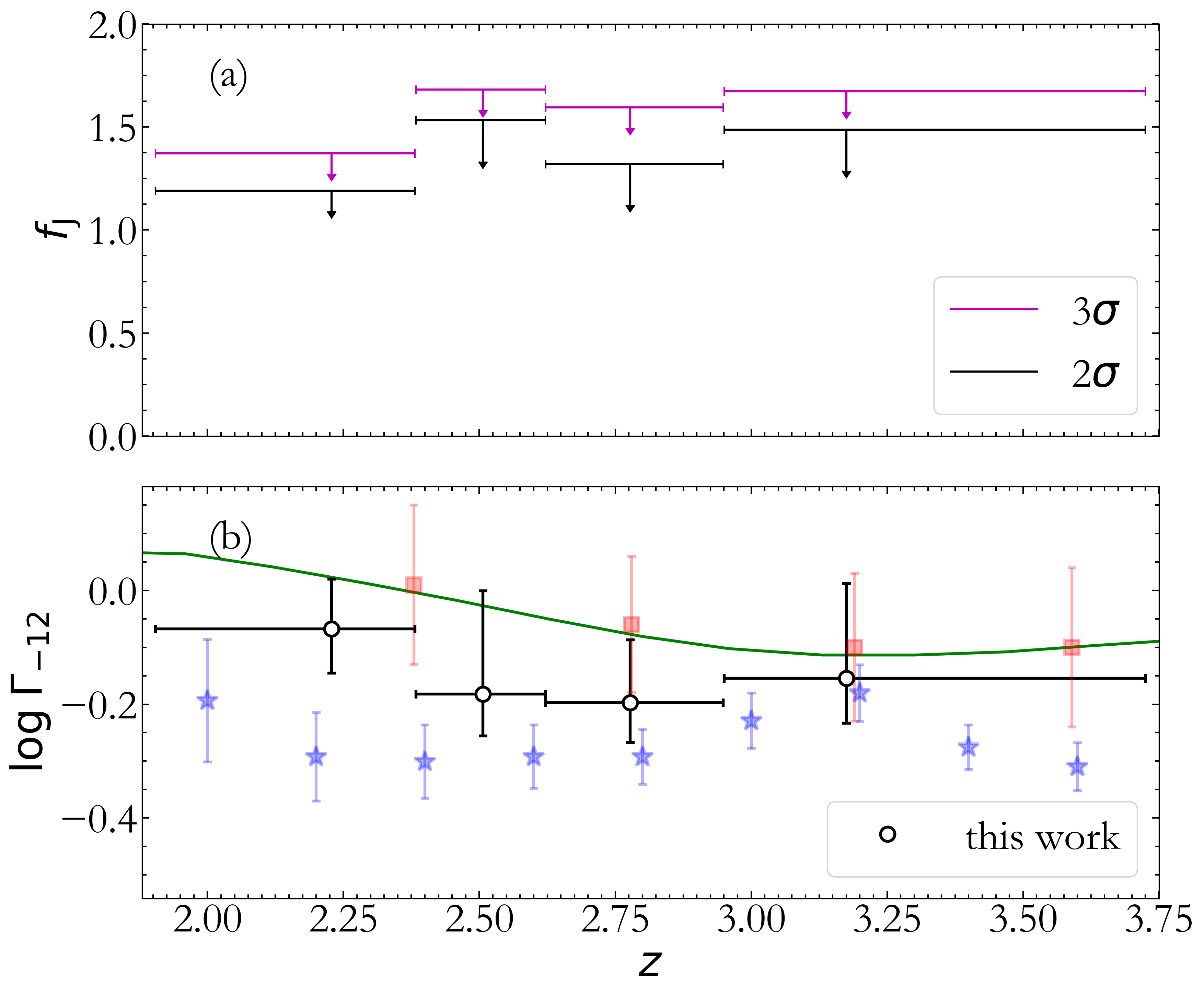}\hspace{1pc}%
\begin{minipage}[b]{13.5pc}\caption{\label{fig:fJ_G12_4}{\sl Top panel:} Upper limits on $f_{\rm J}$ that scales the Jeans smoothing. Black and magenta symbols correspond to the 95.4 and 99.7 per cent one side credible intervals, respectively. {\sl Bottom panel:} Estimation of the hydrogen photoionization rate, $\Gamma_{-12}$, within 4 redshift bins.  Measurements from the present work are shown by black circles, while estimates from \cite{Faucher2008} and \cite{Becker2013MNRAS} assuming $\gamma-1 = 0.63$ and $\gamma-1 = 0.40$ are shown by blue stars and red squares, respectively. Notice that the measurements from \cite{Becker2013MNRAS} are not corrected for the cosmology used in the present paper. The solid curve shows a model from \cite{Khaire2019}.}
\end{minipage}
\end{figure}
The Jeans length in eq~(\ref{eq:bmin}) is \cite{Garzilli2018arXiv}
\begin{equation}\label{eq:lambdaJ}
\lambda_\mathrm{J}= \pi \left(\frac{40}{9}\right)^{1/2} \left(\frac{3\gamma}{5}\right)^{1/2} \left(\frac{k_{\rm B}}{m}\right)^{1/2}\mu^{-1/2} H_0^{-1} (1+z)^{-3/2} \Omega_m^{-1/2} T^{1/2} \Delta^{-1/2},
\end{equation}
where $\Omega_m$ is the matter density parameter, $\mu$ is the mean molecular weight of the gas and $H_0$ is the present-day value of the Hubble parameter.
In a case of a uniform UV background, the column density $N$ can be related to the density $\rho$ as in the model proposed in \cite{Schaye2001}. Taking into account Jeans smoothing and following \cite{Garzilli2015,Garzilli2018arXiv}, we write
\begin{equation}\label{eq:rho-N}
	N = 8.6\times 10^{12}\,\left(\frac{3\gamma}{5}\right)^{1/2} f_{\rm J} \frac{\Delta^{3/2}}{\Gamma_{-12}}\left(\frac{T}{10^4~\mathrm{K}}\right)^{-0.22}\left(\frac{1+z}{3.4}\right)^{9/2}~{\rm cm}^{-2},
\end{equation}
where $\Gamma_{-12}$ is the hydrogen photoionization rate in units of $10^{-12}$~s$^{-1}$. The additional factor $\left({3\gamma}/{5}\right)^{1/2}$\ in eqs~(\ref{eq:lambdaJ}) and (\ref{eq:rho-N}) as compared with \cite{Garzilli2015,Garzilli2018arXiv} accounts
for the non-adiabatic gas behaviour \cite{HuiGnedin1997ApJ}. 
Eqs~(\ref{eq:lambdaJ}) and (\ref{eq:rho-N}) allow to relate the position of the minimal line width in $b-N$ plane, eq~(\ref{eq:bmin}), with the parameters of the effective EOS.
To construct the model for obtained data sample, we write the likelihood function based on eq~(\ref{eq:pdf_N_b}) and take into account the presence of outliers as we did in \cite{Telikova2018b}. The model parameters are $\gamma,\,T_0,\,\Gamma,\,f_{\rm J}$ and nuisance parameters $p,\,\beta$ and a parameter, which characterises a fraction of outliers, thus 7 parameters in total\footnote{Notice, that in \cite{Telikova2018b} we shared the nuisance parameters between the bins. Further analysis have shown that this can lead to the systematic shift in the fit results. Therefore in the present work we discard this sharing.}.
The parameters $\gamma,\,T_0,\,\Gamma$ and $f_{\rm J}$ are strongly correlated \cite{Garzilli2018arXiv} which complicates their measurements. To reduce the uncertainty, we use an additional constraint  based on the measurements of the effective optical depth of the Ly$\alpha$ forest, $\tau_\mathrm{eff}(z)$. This quantity is inferred from the mean transmission of the Ly$\alpha$ forest averaged over many quasars spectra, see \cite{Faucher2008tau}. The effective optical depth can be expressed via the local optical depth $\tau$ and the gas probability density distribution $P(\Delta,z)$ as \cite{Faucher2008}
\begin{equation}
    \tau_{\rm eff}(z) = -\ln\left[\int^\infty_0 P(\Delta,\,z)\tau(z) {\rm d}\Delta \right].
\end{equation}
Following \cite{Faucher2008}, we use the analytical function for the gas probability density distribution, taken from \cite{Miralda-Escude2000}.  
In principle, the local optical depth $\tau(z)$ depends on the spatial extent of an absorber, as discussed above \cite{Garzilli2018arXiv}. However, when the averaging of the different lines of sight is performed, the spatial structures are smeared out and the local Gunn-Peterson approximation \cite{Gunn&Peterson1965Apj,Faucher2008} is applicable.

In our calculations we assumed a standard $\Lambda$CDM cosmology with matter, dark energy and baryon density parameters $\Omega_m = 0.28$, $\Omega_\Lambda = 0.72$ and $\Omega_b = 0.046$, respectively, and 
$H_0=70$~km~s$^{-1}$~Mpc$^{-1}$ \cite{Hinshaw2013ApJS}.

\begin{figure}[t!]
\centering
\includegraphics [width=0.8\textwidth]{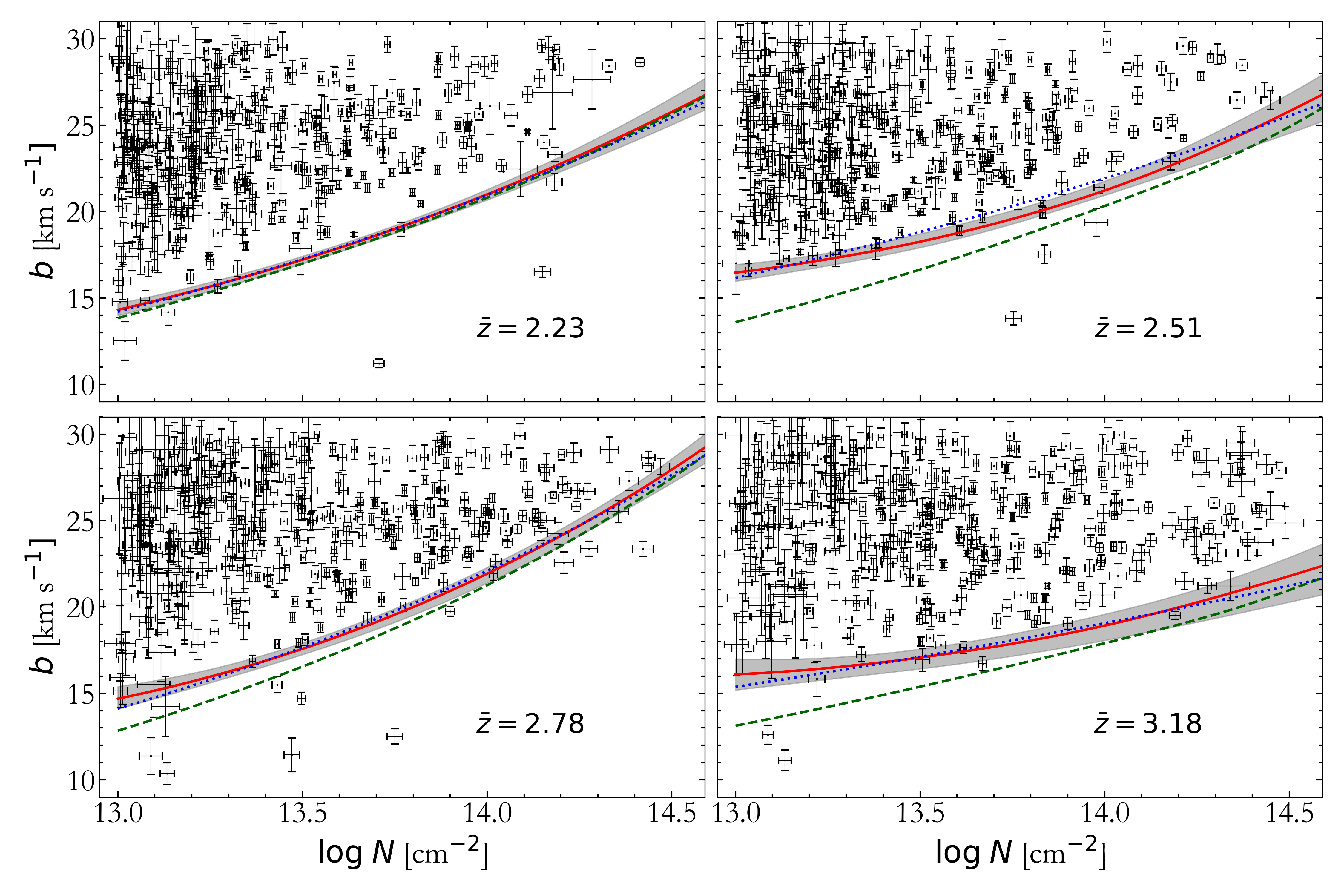}
\caption{Distributions of $b-N$ parameters of Ly$\alpha$ lines in four redshift bins; corresponding mean redshifts are indicated in each panel. Error crosses indicate the measured sample of the Ly$\alpha$ lines from 47 high-resolution quasar spectra. Solid red curves show the lower envelope (with corresponding 68 per cent credible intervals shown by grey areas) of the obtained $b-N$ distribution taking into account nonzero Jeans smoothing of the absorption lines.   Contributions of the pure thermal broadening corresponding to the 50 per cent quantiles are shown by green dashed lines. Blue dotted lines demonstrate the lower envelope of the $b-N$ distribution neglecting Jeans smoothing.}
\label{fig:data_fit_4}
\end{figure}

\section{Results}

We split our data into 4 redshift bins with nearly the same number of absorption lines ($375$ lines in each bin) and estimated parameters in question using the Bayesian framework with the affine Markov Chain Monte Carlo (MCMC) sampler \textsf{emcee} \cite{Foreman-Mackey2013}. Flat priors on the parameters were used. The fit summary is given in table~\ref{tab:pars}. Reported uncertainties correspond to the 68 per cent highest probability density intervals. Dependencies of the $\gamma-1$, $T_0$ and $\Gamma$ on $z$  are shown in the upper and bottom panels in figure~\ref{fig:gamma_T0_4} and in the bottom panel in figure~\ref{fig:fJ_G12_4}, respectively, and compared with measurements by other authors. For the scale parameter $f_{\rm J}$ we were able to estimate only the upper limits as shown in the top panel in figure~\ref{fig:fJ_G12_4}. Estimated cutoffs $b_{\rm min}(N)$ of the $(b-N)$ distributionы are shown for four redshift bins by solid lines in figure~\ref{fig:data_fit_4}. Grey areas correspond to the 68 per cent intervals for $b_{\rm min}(N)$ obtained from the MCMC chain. Contribution of the thermal broadening to the total $b_{\rm min}$ is indicated by the dashed green line. By the dotted blue line in figure~\ref{fig:data_fit_4} we show the cutoff of the distribution, as measured neglecting the Jeans smoothing, i.e. assuming $b_{\rm min}=b_{\rm th}$. A comparison between parameters $\gamma$ in case of nonzero and zero Jeans smoothing contribution is presented in last two columns in table~\ref{tab:pars}.
\begin{figure}[t!]
\centering
\includegraphics [width=0.64\textwidth]{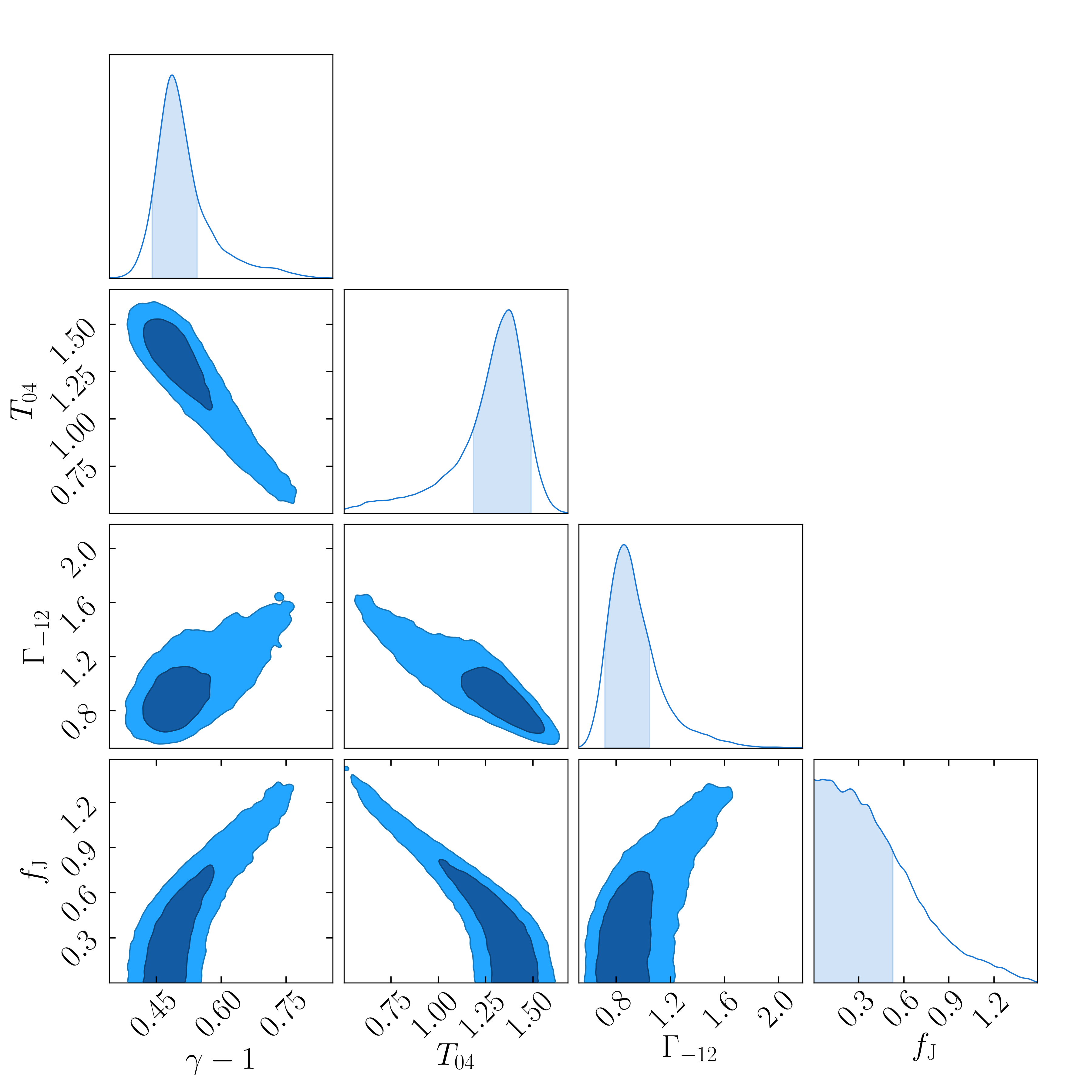}
\caption{Example of the 1D and 2D marginalized posterior distributions of the fit parameters for the redshift bin $z=1.90-2.38$. $T_{04}$ is the temperature $T_0$ measured in units of $10^4$~K.

Dark and light filled areas correspond to the 68 and 95 per cent credible regions, respectively.}
\label{fig:triangle}
\end{figure}

\section{Discussion and conclusions}
Using the technique of the $(b-N)$ distribution analysis, developed in our previous works \cite{Telikova2018a,Telikova2018b,Telikova2019}, we constrained evolution of the parameters of the IGM EOS taking into account the Jeans smoothing contribution to the minimal width of an absorption line. To reduce an impact of the correlations between the parameters, we impose additional constraints based on the measurements of the effective optical depth from \cite{Faucher2008}. An example of the marginalized posterior distributions of the parameters with evident strong correlations between  $\gamma-1$, $T_0$ and $\Gamma$ is shown in figure~\ref{fig:triangle}. 
Although we used additional constraints,  the correlations between the parameters are still sizeable.
At present, we can give only the upper limits on the Jeans smoothing parameter $f_{\rm J}$, which is found to be $\lesssim 1.7$ in all cases.  
To make more certain conclusions about the significance of the Jeans smoothing contribution, it is required to impose additional restrictions on the physical sizes of the absorbers. 
 As seen in figure~\ref{fig:data_fit_4}, dashed lines, the thermal contribution in our model is primarily constrained by 
 high column density regions, where the number of absorbers is relatively small. This is in contrast with the pure thermal model, $b_\mathrm{min}=b_\mathrm{th}$, where the densest regions of the $b-N$ plane (i.e. with small $N$ values, see dotted lines in figure~\ref{fig:data_fit_4}) determine $b_\mathrm{th}$. That means that the neglect of the Jeans smoothing contribution may lead to underestimation of the EOS power-law index $\gamma$ (table~\ref{tab:pars}, see also bottom left panel in figure \ref{fig:triangle}). Moreover we do not find a statistically significant evolution of the EOS parameters with $z$, see figures \ref{fig:gamma_T0_4} and \ref{fig:fJ_G12_4}. This also differs from the case when the pure thermal broadening is assumed, e.g. \cite{Telikova2018b}. Notably, we do not obtain inverted EOS ($\gamma<1$), see table~\ref{tab:pars} and figure~\ref{fig:gamma_T0_4}, top panel. 
 We conclude that the inference of the EOS parameters from the $b-N$ distribution is influenced by the  
 finitude of the IGM filament 
 and this needs to be taken into account \cite{Garzilli2015,Garzilli2018arXiv}. Unfortunately, the correlation between the Jeans smoothing parameter and the parameters of the effective EOS at present do not allow to track their evolution, although the results are in agreement within uncertainties with previous studies.
 One expects much better constraints with an increase of the sample size, i.e. the number of the lines of sight probed in high resolution spectra of the quasars, especially in the high-$N$ region.
\ack 
The work was supported by the Russian Science Foundation, grant 18-72-00110. 
\vspace{-5pt}
\section*{References}
\bibliographystyle{iopart-num}

\providecommand{\newblock}{}

\end{document}